\documentclass[aip,pop,reprint,linenumber,superscriptaddress,amsmath,amsfonts]{revtex4-1}
\usepackage{color,graphicx} 
\usepackage{multirow}

\usepackage[colorlinks,linkcolor=blue,citecolor=blue,bookmarks,pdfstartview=FitH]{hyperref}
\newcommand{\myfigure}{.}

\begin{document}
	\title{Development and testing of an unstructured mesh method for whole plasma gyrokinetic simulations in realistic tokamak geometry}
	\author{Z.X. Lu}
	\author{Ph. Lauber}
	\author{T. Hayward-Schneider}
	\author{A. Bottino}
	\author{M. Hoelzl}
\affiliation{Max-Planck-Institut f\"ur Plasmaphysik,  85748 Garching, Germany}

\begin{abstract}
	In this work, we have formulated and implemented a mixed unstructured mesh-based finite element (FE)–Fourier decomposition scheme for gyrokinetic simulations in realistic tokamak geometry. An efficient particle positioning (particle-triangle mapping) scheme for the charge deposition and field scattering using an intermediate grid as the search index for triangles has been implemented and a significant speed-up by a factor of $\sim30$ is observed as compared with the brute force scheme for a medium-size simulation. The TRIMEG (TRIangular MEsh based Gyrokinetic) code has been developed. As an application, the ion temperature gradient (ITG) mode is simulated using the simplified gyrokinetic Vlasov-Poisson model. Our simulation and that using the ORB5 code for the DIII-D Cyclone case show reasonable agreement. As an additional application, ITG simulations using an ASDEX Upgrade equilibrium have been performed with density and temperature gradient profiles similar to the Cyclone case. Capabilities of the TRIMEG code for simulations with realistic experimental equilibria in the plasma core and in the whole plasma volume with open field lines are demonstrated. 
	

\end{abstract}
\maketitle

\section{Introduction}
Gyrokinetic simulations play an important role in predicting the transport level due to neoclassical physics or turbulence \cite{lee1983gyrokinetic,lin1995gyrokinetic,lin1998turbulent,wang2015distinct,ku2009full,jenko2000electron,dorland2000electron}. One of the leading methods is the particle-in-cell (PIC) method. Numerous PIC codes, such as GTC \cite{lin1995gyrokinetic}, GEM \cite{parker2000electromagnetic}, ORB5 \cite{jolliet2007global}, have been developed for simulations in the core of the tokamak plasmas. Edge plasma simulations have attracted significant attention in recent years due to their connection to, e.g., the high confinement regime of tokamak plasmas; the prediction of the divertor heat-flux width of ITER \cite{chang2017gyrokinetic}; edge localized mode (ELM) control \cite{huysmans2007mhd}.  In order to simulate the edge physics, besides comprehensive physics models \cite{qin2006general,scott2006edge}, numerical schemes such as finite element methods for unstructured meshes in XGC \cite{ku2009full,ferraro2010ideal}, GTS \cite{wang2015distinct} and GTC/GTC-X \cite{nishimura2006finite,bao2019global} and multiple patches of structured meshes in JOREK \cite{huysmans2007mhd} have been developed in order to treat the open field line (OFL) region. While whole plasma simulations for neoclassical transport, ELMs and micro-turbulence have been reported and various numerical schemes have been developed for treating the OFL geometry \cite{ku2009full,chang2017gyrokinetic,huysmans2007mhd,nishimura2006finite,de2019kinetic,vugt2019kinetic}, there is still space to understand the features of different schemes, such as the particle-in-Fourier method (cf. [\!\!\citenum{ameres2018stochastic,ohana2017particle}] and references therein), and thus to optimize the efficiency and the fidelity of the whole volume simulation. 

In this work, we developed the mixed unstructured mesh based finite element–Fourier decomposition scheme, i.e., the mixed particle-in-cell-particle-in-Fourier (PIC-PIF) scheme, for gyrokinetic simulations in general tokamak geometry, including the OFL region. In addition, an efficient particle positioning scheme for the charge deposition and the field gathering using an intermediate grid as the search index for triangles has been implemented. This work is organized as follows. In Section \ref{sec:models}, the physics model and numerical schemes are given. In Section \ref{sec:results}, we perform convergence/scaling studies and simulations of ion temperature gradient (ITG) modes using (I) the concentric circular magnetic geometry and the DIII-D cyclone parameters and (II) the AUG realistic magnetic equilibrium and analytical density and temperature profiles. In \ref{sec:conclusion}, conclusions and an outlook are given. 

\section{Physics model and numerical methods}
\label{sec:models}
In the following, we will define the normalization in Section \ref{subsec:normalization}. In Sections \ref{subsec:equilibrium}--\ref{subsec:field}, we will describe the three basic classes in the code, namely, the equilibrium,  particle and field classes. We then describe the numerical methods in Section \ref{subsec:numerical_methods}. 
While the equations of motion for the guiding center, the weight equation, and the gyrokinetic Poisson equation (in the long wavelength limit) in this work are the same as or are the simplified version of other codes such as ORB5 \cite{jolliet2009gyrokinetic,lanti2019orb5arxiv}, the mixed unstructured mesh-based finite element-Fourier decomposition scheme makes our work different from ORB5. In ORB5, the OFL region is not included in either the particle pusher or the field solver. The mixed approach in this work thus also serves as a potential candidate for the extension of the present codes such as ORB5 towards whole plasma volume simulations.
The details related to the finite element and unstructured meshes are described in another work for circular tokamak geometry \cite{lu2019mixed} and will be omitted in this work. 

\subsection{Normalization}
\label{subsec:normalization}
Normalization units are defined and physics quantities are normalized to the normalization units. The length unit is $R_N=1m$. The  velocity unit is $v_N=v_{th,hy}$, where $v_{th,hy}=\sqrt{2T_{hy}/m_{hy}}$, $T_{hy}$ is the reference temperature, $m_{hy}$ is the the mass of hydrogen, the subscripts `$N$' and `$hy$' indicate ``normalization'' and ``hydrogen'' respectively.  For each particle species s, $v_{th,s}$ is used as the velocity unit for the particle initialization according to a Maxwellian distribution function while unit conversion to the normalization unit $R_N$ is performed in the equations of motion for the guiding center and the field equation. The time unit is $R_N/v_{th,hy}$. The magnetic field unit is $B_N=1T$. 

\subsection{The coordinates and the equilibrium}
\label{subsec:equilibrium}
In the right-handed coordinates $(R,\varphi,Z)$ and $(\psi,\varphi,\theta)$, where $\psi$ is the poloidal flux function, using the EFIT convention, the magnetic field is represented as  
\begin{equation}
    {\bf B}=\nabla\psi\times\nabla\varphi + F\nabla\varphi\;\;, 
\end{equation}
where $F$ is the poloidal current function. 
In the $(\psi,\varphi,\theta)$ coordinates, the safety factor is defined as $q={{\bf B}\cdot\nabla\varphi}/({{\bf B}\cdot\nabla\theta})={JF}/{R^2}$, 
where $J=\{\nabla\psi\times\nabla\varphi\cdot\nabla\theta\}^{-1}$. 
The equilibrium variables are constructed using B-splines in the $(R,Z)$ plane of the $(R,\varphi,Z)$ coordinates. Equilibrium variables such as $B$, $B_R$, $B_Z$ and their derivatives in $R$ and $Z$ directions can be obtained using the B-spline subroutines. 

\begin{figure}\centering
	\includegraphics[width=0.22\textwidth]{\myfigure/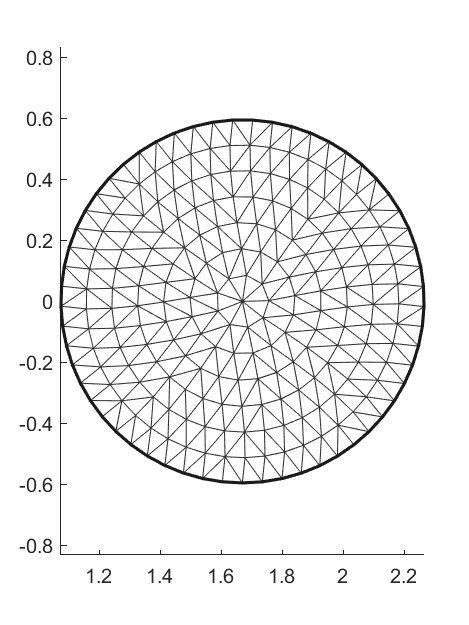}
	\includegraphics[width=0.22\textwidth]{\myfigure/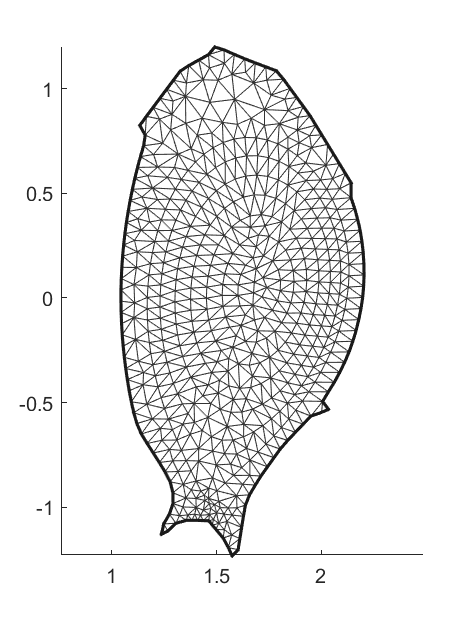}
	\caption{The grids and meshes for the Cylone case and the AUG case used in Section \ref{sec:results}. The grids shown in the figure are much sparser than those used in simulations. }   
	\label{fig:meshes}
\end{figure}

The hybrid coordinates $(R,\varphi,Z)$ and $(\psi,\varphi,\theta)$ are used. On the one hand, the user specified computation grids are aligned along the magnetic flux surface using the $\psi$ coordinate, or, if the OFL region is included, along the plasma boundary; the refinement grids are generated when the Delaunay refinement algorithm is called for the generation of the unstructured meshes \cite{shewchuk2002delaunay}. The refinement grids using the Delaunay algorithm are not necessarily along the magnetic flux surface but it is a widely used technique for the improvement of the mesh quality. Two cases of the grids and the unstructured meshes for the simulations in Section \ref{sec:results} are shown in Fig. \ref{fig:meshes}. More details about the finite element for the unstructured meshes are in another work \cite{lu2019mixed}. On the other hand, in order to treat the open field line geometry, the equations for the field and particles are solved in $(R,\varphi,Z)$ coordinates. 
In $(R,\varphi,Z)$ coordinates, the equilibrium magnetic field is expressed as 
\begin{equation}
    B_R=-\frac{1}{R}\frac{\partial\psi}{\partial Z}\;\;,\;\;
    B_Z=+\frac{1}{R}\frac{\partial\psi}{\partial R}\;\;,\;\;
    B_\varphi=\frac{F}{R}\;\;.
\end{equation}
TRIMEG relies on the B-Spline subroutines for the interpolation of $B_R$ and $B_Z$ accurately and keeps the error in $\nabla\cdot{\bf B}$ at a low level. For simulations in this manuscript, we use $N_R=N_Z=129$ as the grid numbers in R and Z directions for the equilibrium construction. In the relevant simulation domain, the error in $\nabla\cdot{\bf B}/B$ is well below $10^{-4}$ with the upper limit of $10^{-3}$. 
The parallel derivative is
\begin{eqnarray}\label{eq:parallel_derive_RZ}
    \partial_\parallel\delta\phi=\left(b_R\partial_R+b_Z\partial_Z+\frac{b_\phi}{R}\partial_\phi\right)\delta\phi \;\;, 
\end{eqnarray}
where $\delta\phi$ is the perturbed electrostatic scalar potential, $b_R=B_R/B$, $b_Z=B_Z/B$ and $b_\phi=B_\phi/B$. 

The field-aligned coordinates are constructed as another option for the calculation of the parallel derivatives in addition to Eq. \ref{eq:parallel_derive_RZ}. Along the magnetic field line $\bf B$, the auxiliary Clebsch coordinates $(\chi,\xi,l)$ are determined by 
\begin{eqnarray}
	\left.\frac{dR(\chi,\xi,l)}{d\phi}\right|_{\chi,\xi}&=&R\frac{B_R}{B_\phi}\;\;, \\
	\left.\frac{dZ(\chi,\xi,l)}{d\phi}\right|_{\chi,\xi}&=&R\frac{B_Z}{B_\phi}\;\;,\\
	\left.\frac{dl}{d\phi}\right|_{\chi,\xi}&=&R\frac{B}{B_\phi} , 
\end{eqnarray}
where $(\chi,\xi)$ labels a magnetic field line and can be taken as $(R,Z)$ at $\varphi=0$ for convenience and $l$ is the coordinate along $\bf B$. The parallel derivative using these Clebsch coordinates is
\begin{equation}\label{eq:parallel_derive_clebsch}
    \partial_\parallel\delta\phi=\left.\frac{\partial}{\partial l}\right|_{\chi,\xi}\delta\phi \;\;.
\end{equation}
For parallel derivatives of high $n$ (toroidal mode number) field-aligned modes, Eq. \ref{eq:parallel_derive_clebsch} can provide high accuracy compared with Eq. \ref{eq:parallel_derive_RZ}. 

Since $(R,\varphi,Z)$ or $(\chi,\xi,l)$ coordinates are used for solving all equations in the code except when the grids are initialized according to $\psi$, and the safety factor defined in $(\psi,\varphi,\theta)$ does not appear explicitly but $B_R$, $B_Z$ and $B_\varphi$ are used directly in our equations, the singularity of the $q$ at the X point does not appear in any equation solved in TRIMEG. The auxiliary coordinates $(\chi,\xi,l)$ maintain the high accuracy of the parallel derivative calculation for high $n$ field-aligned modes and are similar to the flux-coordinate independent approach \cite{hariri2013flux,stegmeir2016field}. Discussions related to the flux coordinates and singularity of the safety factor at the X point can be found in [\!\!\citenum{lu2012theoretical}] and references therein. 

\subsection{Particles}
\label{subsec:particles}
\subsubsection{Equations of motion}
In this work, we focus on electrostatic simulations. In order to describe the particle guiding center motion, we follow the canonical Hamiltonian equations \cite{white1990canonical}. Using $({\bf {R}},\rho_\parallel,\mu)$ as the guiding center coordinates, the equations of motion are as follows,
\begin{eqnarray}
    \dot{\bf R}&=&\frac{1}{Ze{\bf B}\cdot{\bf B_\parallel}^*}
    \left[\frac{Z^2e^2B^2\rho_\parallel}{m}{\bf B_\parallel}^*
    +{\bf B}\times\nabla H\right]\;\;, \\
    \dot{\rho}_\parallel &=& -\frac{1}{Ze{\bf B}\cdot{\bf B_\parallel}^*}
    \left({\bf B}_\parallel^*\cdot\nabla H \right)\;\;,
\end{eqnarray}
where ${\bf B}_\parallel^*={\bf B}+\rho_\parallel\nabla\times{\bf B}$, $\rho_\parallel=mv_\parallel/(ZeB)$, $Z$ is the charge number, $H=Z^2e^2\rho_\parallel^2B^2/(2m)+m\mu B+Ze\delta\phi$ is the Hamiltonian, and $\mu=v_\perp^2/(2B)$. 
The equations of motion above are equivalent to those adopted in XGC \cite{ku2009full}, 
\begin{eqnarray}
    {\bf \dot x}&=&(1/D)[v_\parallel{\bf b}+(v^2_\parallel/B) \nabla B\times{\bf b}   \nonumber \\
    &+&  B\times(\mu\nabla B-E)/B^2]\;\;\\
    \dot v_\parallel &=& -(1/D)({\bf{B}}+v_\parallel\nabla B\times{\bf b})\cdot(\mu\nabla B-E),
    \end{eqnarray}
where $D\equiv 1+v_\parallel {\bf b}\cdot \nabla\times{\bf b}/B$ is related to the higher order corrections.

The variables ${\bf \dot{R}},\dot{v}_\parallel$ can be written as 
\begin{eqnarray}
	{\bf \dot{R}}&=&{\bf v}_{\parallel 0}+{\bf v}_{d0}+{\bf v}_E \;\;, \\
	\dot{v}_\parallel&=&\dot{v}_{\parallel 0}+\dot{v}_{\parallel E} \;\;,
\end{eqnarray}
where the subscripts $0$ and $E$ indicate the motion of the guiding center in equilibrium and that due to the electrostatic field. In $(R,\varphi,Z)$ coordinates, the contravariant components of the velocity $v^\alpha\equiv{\bf v}\cdot\nabla\alpha$ are calculated for different terms. 
\begin{enumerate}
	\item Magnetic drift ${\bf v}_{d0}$
	\begin{eqnarray}
	v_{d0}^R &=& \frac{v_\parallel^2+v_\perp^2/2}{\omega_c}\frac{b_\varphi}{B_\parallel^*} \partial_ZB \;\;, \\
	v_{d0}^Z &=& -\frac{v_\parallel^2+v_\perp^2/2}{\omega_c}\frac{b_\varphi}{B_\parallel^*} \partial_RB \;\;, \\
	v_{d0}^\varphi &=& \frac{v_\parallel^2+v_\perp^2/2}{\omega_c}\frac{1}{RB_\parallel^*} (-b_R\partial_ZB+b_Z\partial_RB) \;\;. 
	\end{eqnarray}
	The dominant terms of the equations of motion are obtained by omitting the terms of the order of $\rho*=\rho_{th}/a$ and of $B_\theta/B$, where $\rho_{th}=mv_{th}/(ZeB)$, $v_{th}=\sqrt{2T/m}$. In $(R,\varphi,Z)$ coordinates, 
	the dominant terms in the equations of motion are
	\[
	{\bf v}_{d0}\approx\frac{v_\perp^2+2v_\parallel^2}{2\omega_cB} {\bf b}\times\nabla B\;\;,
	\]
	which, noticing that ${\bf b}\approx b_\phi\hat\phi$, where $\hat\phi$ is the unit vector in the toroidal direction, and $B\approx B_\phi$, can be further reduced to,
	\begin{eqnarray}
	\label{eq:vd_simple}
	v_{d0}^R\approx0\;\;,\;\;
	v_{d0}^Z\approx-b_\phi\frac{v_\parallel^2+v_\perp^2/2}{\omega_cB} \partial_R B_\phi\;\;.
	\end{eqnarray}
	
	\item $E\times B$ drift ${\bf v}_E$. Generally, $E\times B$ drift ${\bf v}_E$ is contributed from the equilibrium scalar potential $\delta\phi_{eq}$ and the perturbed scalar potential $\delta\phi$, i.e.,  ${\bf v}_E={\bf v}_{E,eq}+\delta{\bf v}_E$. In this work, we only consider the latter one, i.e., 
	\begin{eqnarray}
	\delta v_E^R &=&\frac{1}{B_\parallel^*}(b_\varphi\partial_Z\langle\delta\phi\rangle-\frac{b_Z}{R}\partial_\varphi\langle\delta\phi\rangle) \;\;,\\
	\delta v_E^Z &=&\frac{1}{B_\parallel^*}(-b_\varphi\partial_R\langle\delta\phi\rangle+\frac{b_R}{R}\partial_\varphi\langle\delta\phi\rangle) \;\;, \\
	\delta v_E^\varphi &=&\frac{1}{RB_\parallel^*}(-b_R\partial_Z\langle\delta\phi\rangle+b_Z\partial_R\langle\delta\phi\rangle) \;\;,
	\end{eqnarray}
	where $\langle\ldots\rangle$ indicates a gyro average. In TRIMEG, the four point gyro average scheme is adopted. For the calculation of $\langle A(R,Z)\rangle$, where $A$ indicates $\delta\phi$ or its derivative in $R$ or $Z$ direction, $A(R\pm\rho_\perp,Z)$ and $A(R,Z\pm\rho_\perp)$ are calculated and the average value is obtained. 
	The dominant term is
	\begin{eqnarray}
	\label{eq:vE_simple}
	\delta {\bf v}_{E}=\frac{{\bf b}\times\langle\nabla\delta\varphi\rangle}{B} = \frac{1}{B}\langle[\hat{R}\partial_Z\delta\varphi - \hat{Z}\partial_R\delta\varphi]\rangle\;\;,\;\;
	\end{eqnarray} 
	where $\hat R$ and $\hat Z$ are the unit vectors in the $R$ and $Z$ directions.
	\item Parallel acceleration $\dot{v}_{\parallel0}$
	\begin{eqnarray}
	\dot{v}_{\parallel0} = -\frac{\mu}{B_{\parallel}^*}(B_R^*\partial_RB+B_Z^*\partial_ZB).
	\end{eqnarray}
	The dominant term is 
	\begin{eqnarray}
	\label{eq:dvpardt_simple}
	\dot{v}_{\parallel0} \approx -{\mu}(b_R\partial_RB+b_Z\partial_ZB).
	\end{eqnarray}
	
\end{enumerate}
The simplification of other terms such as $v_\parallel$ and $\dot{v}_{\parallel E}$ is trivial and is omitted.

\subsubsection{Weight equation}\label{subsubsec:weight_eq}
The gyrokinetic Vlasov equation for the evolution of the perturbed distribution function $\delta f({\bf R},v_\parallel,\mu)$ is
\begin{eqnarray} \label{eq:dfdt0}
	\frac{d\delta f}{dt}=\tau({\bf E})\;\; ,
\end{eqnarray} 
where 
\[
\tau({\bf E})=-f_0 \delta {\bf R}\cdot\nabla \ln f_0
+ f_0\frac{e}{m} \langle {\bf E} \rangle \cdot  \left.\frac{d{\bf R}}{dt}\right|_0 \;\;.
\]

The weight of the markers is defined to represent the perturbed distribution function, 
\begin{eqnarray} \label{eq:delta_w}
	\delta f({\bf R},v_\parallel,\mu) &=& \frac{N_{ph}}{N}\sum_{p=1}^N\frac{1}{2\pi B_\parallel^*} w_p(t)\delta({\bf R}-{\bf R}_p(t))  \nonumber\\ &&\times\delta(v_\parallel-v_{\parallel,p}(t))\delta(\mu-\mu_p(t))\;\;,
\end{eqnarray}
where $N_{ph}$ and $N$ are the total numbers of the physical particles and numerical markers respectively and the subscript $p$ is the marker index. 
Defining $\Omega_p$ as the phase space volume occupied by the marker $p$,  Eq. \ref{eq:delta_w} yields
\begin{eqnarray}
	\delta f({\bf R}_p,v_{\parallel,p},\mu_p)\Omega_p = \frac{N_{ph}}{N}w_p(t)\;\;,
\end{eqnarray} 
and Eq. \ref{eq:dfdt0} gives
\begin{eqnarray} \label{eq:dwdt0}
	\frac{dw_p}{dt} = \frac{N}{N_{ph}}\Omega_p\tau({\bf E})\;\;.
\end{eqnarray}
The above definition is the same as that in ORB5 \cite{jolliet2007global,lanti2019orb5arxiv}. However, in this work, for the sake of simplicity, we load the markers with the distribution function the same as that of the physical particles, i.e.,
\begin{eqnarray} \label{eq:fmk_fph}
	f_\text{mark}=\frac{N}{N_{ph}} f\;\;,
\end{eqnarray}
then 
\begin{eqnarray}\label{eq:omegap}
	\Omega_p\equiv\frac{B_\parallel^*d{\bf R}d\mu d\alpha}{dN}=\frac{1}{f_\text{mark}} \;\;,
\end{eqnarray}
where $\alpha$ is the gyro angle.
Equations \ref{eq:dwdt0}, \ref{eq:fmk_fph} and \ref{eq:omegap} yield
\begin{eqnarray}
	\frac{dw_p}{dt} = - \delta {\bf R}\cdot\nabla \ln f_0
	+ \frac{e}{m} \langle {\bf E} \rangle \cdot  \left.\frac{d{\bf R}}{dt}\right|_0 \;\;.
\end{eqnarray}

The perturbed density in a small volume $\Delta V$ is calculated from the marker weight in $\Delta V$
\begin{equation}
    \frac{\delta n}{\langle n\rangle}
    =\frac{V_{tot}}{N\Delta V}\sum_{p\in\Delta V}w_p \;\;,
\end{equation} 
where $\langle n\rangle$ is the volume averaged density. For unstructured meshes, the volume $\Delta V$ is centered around a vertex and $\Delta V=\Delta V_\text{vert}$ is calculated using $\Delta V_\text{tria}$, where $\Delta V_\text{tria}$ is a triangular prism which extends along the $\varphi$ direction. Using the particle-in-Fourier method in the toroidal direction, we have for each toroidal mode number $n$,
\begin{eqnarray}
\label{eq:deltan_n_weight}
\frac{\delta n_n(R,Z)}{\langle n\rangle}=\frac{V_{tot}}{N\Delta V} 
\sum_{(R_p,Z_p)\in\Delta S}w_p e^{-in\varphi_p}\;\;,
\end{eqnarray}
where $\Delta S$ is the projection of $\Delta V$ in the $(R,Z)$ plane. 

\subsection{Field equation}
\label{subsec:field}
The gyrokinetic Poisson equation with the long wavelength approximation is adopted in this work, i.e., 
\begin{equation}\label{eq:poisson0}
    -\nabla_\perp\frac{n_0}{\omega_c B}\cdot\nabla_\perp\delta\phi=\delta n_i-\delta n_e \;\;.
\end{equation}
Generally, the electron response is dominated by the adiabatic response. Thus, the electron response can be decomposed into the adiabatic and non adiabatic (NA) parts, i.e., 
\begin{equation}\label{eq:deltane0}
    \delta n_e=\frac{e}{T_e}\delta\tilde{\phi} + \delta n_e^{NA}\;\;,
\end{equation}
where $\delta\tilde{\phi}$ is the non-zonal component, i.e., $\delta\tilde{\phi}=\delta{\phi}-\delta{\phi}_{0,0}$, $\delta\phi_{0,0}$ is the poloidal harmonic with $n=0,m=0$, where $m$ is the poloidal mode number. Notice that the Fourier decomposition is used in the $\varphi$ direction but the finite element method is used in the $(R,Z)$ plane. 

For $n\neq0$, with the subscript $n$ omitted, 
\begin{equation} \label{eq:poisson_na0}
    -\nabla_\perp\frac{n_0}{\omega_c B}\cdot\nabla_\perp\delta\phi+\frac{e}{T_e}\delta\tilde{\phi} =\delta n_i-\delta n_e^{NA} \;\;.
\end{equation}

For $n=0$, 
\begin{equation} \label{eq:poisson_na1}
    -\nabla_\perp\frac{n_0}{\omega_c B}\cdot\nabla_\perp\delta\phi
    +\frac{e}{T_e}(\delta\tilde{\phi}-\delta\phi_{0,0}) =\delta n_i-\delta n_e^{NA} \;\;.
\end{equation}
In this work, we focus on the $n\neq0$ modes while the studies involving $n=0$ components such as the geodesic acoustic mode will be reported in another separate work \cite{lu2019mixed}. 

\subsection{Numerical methods}
\label{subsec:numerical_methods}
\subsubsection{General description}
This gyrokinetic Poisson-Vlasov system is implemented in Fortran. The field equation is solved using the finite element method for unstructured meshes. The sparse matrix corresponding to the gyrokinetic Poisson equation is solved using PETSc (Portable, Extensible Toolkit for Scientific Computation) \cite{balay2019petsc}. The Runge-Kutta fourth order integrator is implemented for particles and coupled to the field solver. 
The Runge–Kutta fourth-order method is given by the following steps, 
\begin{eqnarray}
&&	\Delta {\bf X}_1=\Delta td_t{\bf X}({\bf X}_t,\delta\phi_t)\;\;,\;\;  \;\;\;\;\;
	\bf X_t^{(1)}={\bf X}_t+\Delta{\bf X}_1/2\nonumber\\
&&	\Delta {\bf X}_2=\Delta td_t{\bf X}(\bf X_t^{(1)},\delta\phi_t^{(1)})\;\;,\;\;
	\bf X_t^{(2)}={\bf X}_t+\Delta{\bf X}_2/2 \nonumber\\
&&	\Delta {\bf X}_3=\Delta td_t{\bf X}(\bf X_t^{(2)},\delta\phi_t^{(2)})\;\;,\;\;
	\bf X_t^{(3)}={\bf X}_t+\Delta{\bf X}_3/2\nonumber\\
&&	\Delta {\bf X}_4=\Delta td_t{\bf X}(\bf X_t^{(3)},\delta\phi_t^{(3)})\;\;,\nonumber\\
&&	{\bf X}(t+\Delta t)={\bf X}(t)+(\Delta {\bf X}_1+2\Delta {\bf X}_2+2\Delta {\bf X}_3+\Delta {\bf X}_4)/6\;\;, \nonumber
\end{eqnarray}
where $d_t{\bf X}=d{\bf X}/dt$, ${\bf X}=({\bf R},v_\parallel)$,  $\Delta t$ is the time interval and $\delta\phi^{(i)}$ is solved from Eq. \ref{eq:poisson0} with $\delta n(t)$ obtained using ${\bf X}_t^{(i)}$ for $i=1,2,3$.

\subsubsection{Particle positioning (deposition/gathering) scheme}
\label{subsubsec:particle_positioning}
When calculating the toroidal component of the charge density perturbation $\delta n_n(R,Z)$ in Eq. \ref{eq:deltan_n_weight} using marker weights $w_p$ in the so-called ``charge deposition'' stage, or when interpolating the field value at the particle position using the grid field value during the ``field gathering'' stage, the marker-triangle mapping, i.e., the particle positioning, needs to be treated. For the brute force particle position scheme, each triangle is checked for each marker whether the triangle contains the marker, which leads to a $NN_t$ scale computational cost, where $N$ and $N_t$ are the marker and triangle numbers respectively. In this work, rectangular grids (``boxes'') are constructed in $(R,Z)$ space and the box-triangle index $\{\text{Box}_i \mapsto \text{Triangle}_j\}$ is built when there is overlap between a box $i$ and a triangle $j$.  The mapping $\{\text{Box}_i \mapsto \text{Triangle}_j\}$, $j=1,\ldots,N_\text{i,tri}$  is stored in the dynamically growing arrays for each box $i$. For a given marker $p$, the box which contains Marker $p$ is first found, i.e., the $\{p\mapsto \text{Box}_i\}$ mapping is identified. Then using the box-triangle mapping  $\{\text{Box}_i \mapsto \text{Triangle}_j\}$, the corresponding triangle is identified. The computational cost is $\alpha N$ for $N$ markers, where $\alpha$ is a constant number.

\begin{figure*}\centering
	\includegraphics[width=0.3\textwidth]{\myfigure/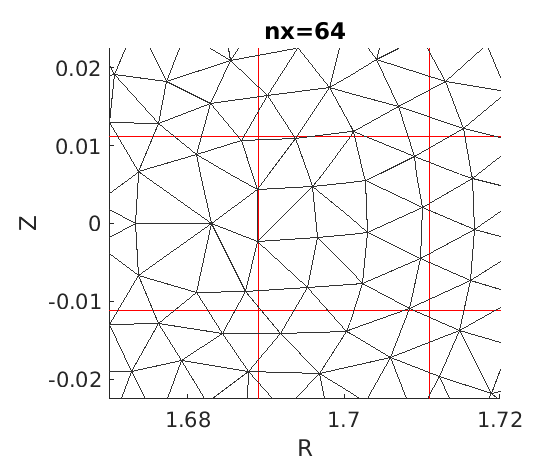}
	\includegraphics[width=0.3\textwidth]{\myfigure/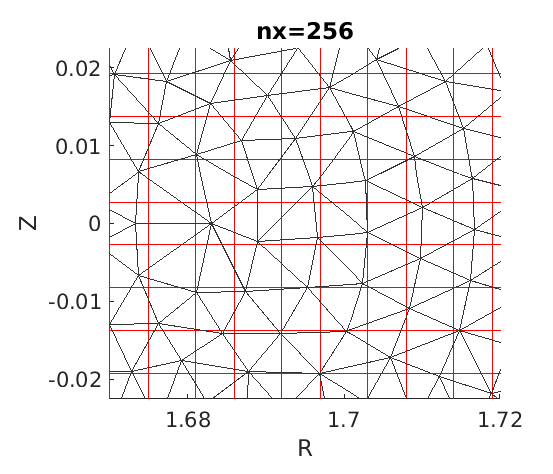}
	\includegraphics[width=0.3\textwidth]{\myfigure/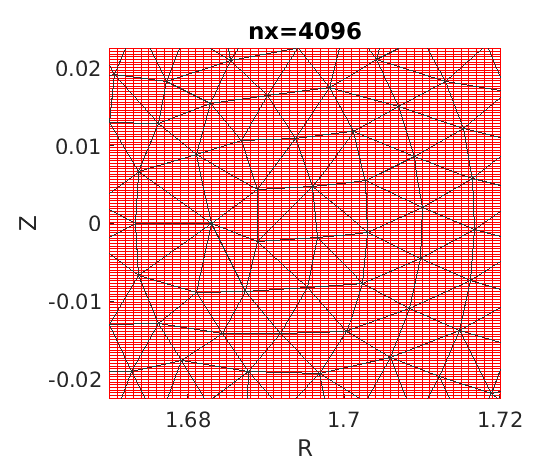}
	\caption{The box (red lines) size and its comparison with the triangle (black lines) size for different values of $N_x$. Cyclone parameters are used and the radial grid number $N_r=90$. }   
	\label{fig:box_triangle}
\end{figure*}

The intermediate boxes are generated in the simulation domain with given $N_x$ and $N_y$, where $N_x$ and $N_y$ are the rectangular grid numbers in $R$ and $Z$ directions respectively. One limit is $N_x\ll N_r$, where $N_r$ is the radial grid number of the unstructured meshes. For $N_x=2$ (the box number is one), the positioning scheme is identical to the brute force scheme.  The other limit is $N_x\gg N_r$, for which the box size is much smaller than the triangle size. A typical case between these two limit cases is that with $N_x\approx N_r$. These three cases are shown in Fig. \ref{fig:box_triangle}. The computational speed-up versus the box size or $N_x$ will be studied in Sec. \ref{sec:results}.

\section{Numerical results}
\label{sec:results}
\subsection{Parameters and simplifications for the simulation}
\label{subsec:parameters}
In this section, two experimental cases are discussed. For numerical studies and benchmarks in Sections \ref{subsec:converge_scaling} and \ref{subsec:itg_cyclone}, the DIII-D Cyclone case is adopted, for which the parameters are the same as those in the benchmark work  \cite{gorler2016intercode}.
The geometry with concentric circular magnetic flux surfaces is assumed. The nominal safety factor profile is\cite{gorler2016intercode} 
\begin{eqnarray}
\label{eq:qprofile0}
q(r)=2.52\bar{r}^2-0.16\bar{r}+0.86\;\;,
\end{eqnarray}where $\bar{r}\equiv r/a$.
In Sections  \ref{subsec:converge_scaling} and \ref{subsec:itg_cyclone}, an ad hoc equilibrium model is adopted. By assuming the following form of the safety factor profile,
\begin{eqnarray}
\bar{q}(r)=\bar{q}_0+\bar{q}_2\bar{r}^2\;\;,
\end{eqnarray}
where $\bar{q}=q/\sqrt{1-r^2/R_0^2}$, the poloidal flux function can be obtained analytically \cite{bottino2007nonlinear,lu2019mixed}. The values of $q$ and the magnetic shear $\hat s$ are matched to Eq. \ref{eq:qprofile0} at $r_c=0.5a$, i.e., $q_2=q_c\hat{s}_c/(2r_c^2)$, $q_0=(2/\hat{s}_c-1)\bar{r}_c^2 q_2$, where $q_c$ and $\hat{s}_c$ are calculated at $r=r_c$ using Eq. \ref{eq:qprofile0}.
The temperature and density profiles indicated by $A(r)$ and the corresponding normalized logarithmic gradients indicated by $L_{ref}/L_A$ are given by
\begin{eqnarray}
\label{eq:nT}
\frac{A(r)}{A(r_0)}&=&\exp\left\{-\kappa_A W_A\frac{a}{L_{ref} }\tanh\left( \frac{r-r_c}{W_A a} \right) \right\}\;\;,\\
\label{eq:dlnnTdr}
\frac{L_{ref}}{L_A}&=&-L_{ref} \frac{d\ln A}{dr}=\kappa_A\cosh^{-2} \left( \frac{r-r_c}{W_A a} \right) \;\;,
\end{eqnarray}
where the subscript `c' denotes the center of the gradient and the values of $r_c$, $W_A$ etc are in Table \ref{tab:parameters}. 

For the studies using the realistic geometry in Section \ref{subsec:itg_aug}, the ASDEX Upgrade (AUG) case with shot number  34924 at 3.600s is chosen. This is a typical discharge for the study of energetic particle and turbulence physics \cite{lauber2018strongly}. In the simulation, we use the experimental equilibrium but use the analytical density and temperature profiles in Eq. \ref{eq:dlnnTdr}, with the radial coordinate replaced with $\rho_p=\sqrt{(\psi-\psi_0)/(\psi_b-\psi_0)}$, where $\psi_0$ and $\psi_b$ are the poloidal magnetic flux function at the magnetic axis and at the last closed surface respectively. The purpose of this study is to test the capability of treating the realistic geometry with minimum technical complexity. The fully self consistent treatment of the density/temperature profile and the equilibrium will be addressed in another work. 

Since our purpose is to study the mixed PIC-PIF scheme and the particle search scheme in this work and address the basic ITG mode problem in the whole plasma geometry with minimum complexity, we have made the following simplifications. 
\begin{enumerate}
	\item Only the dominant terms in the equations of motion, Eqs. \ref{eq:vd_simple}, \ref{eq:vE_simple} and \ref{eq:dvpardt_simple}, are solved.
	\item The equilibrium variation of $n$, $B$ and $T$ in the gyrokinetic Poisson equation, Eq. \ref{eq:poisson0}, is ignored. 
	\item The ITG instability drive in Eq. \ref{eq:dlnnTdr}  for the weight equation is kept but the equilibrium variation in $n$, $T$ and $B$ is  omitted. 
	\item A single toroidal harmonic is simulated without the nonlinear terms, even though the dominant nonlinear term $\delta v_E\cdot\nabla\delta f$ for the ITG saturation is implemented in TRIMEG.
	\item The Dirichlet boundary condition is adopted for the gyrokinetic Poisson equation with $\delta\phi=0$ at the boundary. The ``absorbing boundary condition'' for markers are adopted, i.e., the markers hitting the boundary are removed from the system. 
	\item Adiabatic electron approximation is adopted, i.e., $\delta n_e^{NA}=0$ in Eq. \ref{eq:deltane0}.
	\item As the initial condition, markers with Maxwellian distribution are loaded in the simulation domain. Markers hitting the wall are removed (absorbing boundary condition). Since in this work, we only performed linear simulations, the marker distribution does not change after all absorbed markers are removed. 
\end{enumerate} 
This simplified model can be replaced with a more comprehensive one by either future development of the TRIMEG code, or by implementing the finite element solver for unstructured meshes and the PIC-PIF scheme in other codes such as ORB5 and GTC. 

\begin{table}\begin{center}
		\begin{tabular}{c|c|c|c|c|c|c|c}
			\hline 
			$r_c/a$& $a/R_0$ & $a/\rho_s$ & $T_e/T_i$ & $R_0/L_{ref}$ & $\kappa_{Ti}$ & $\kappa_{n}$ & $W_n=W_{Ti}$ \\ 
			\hline 
			0.5    & 0.36    & 180      & 1         &  1 & 6.69& 2.23 & 0.3\\ 
			\hline 
		\end{tabular} 
	\end{center}\caption{Parametes for ITG with adiabatic electrons (same as those in Ref. \citenum{gorler2016intercode}), where  $R_0=1.67m$, $\rho_s= c_s/\Omega_i$, $c_s=\sqrt{T_e/m_i}$ and the value of $\rho_s$ is calculated using $T_e(r/a=0.5)$ and $B(r=0)$.}\label{tab:parameters}
\end{table}

\subsection{Convergence and scaling studies}
\label{subsec:converge_scaling}
\begin{figure}\centering
	\includegraphics[width=0.48\textwidth]{\myfigure/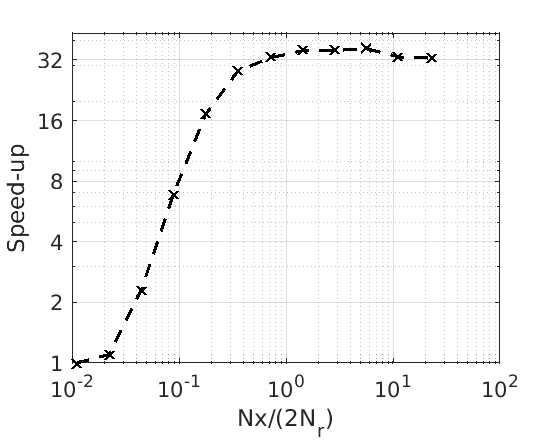}
	\caption{The speed-up of a medium size simulation ($N_r=90$, 25.6 million markers) for different values of $N_x/(2N_r)$ with respect to that with $N_x=2$ (the brute force scheme).}   
	\label{fig:box_size_scan}
\end{figure}

The effects of the rectangular box size on the computational cost in the particle positioning scheme are studied using a medium size case whose radial grid number is $N_r=90$ and the total marker number $N$ is 25.6 million. The brute force scheme ($N_x=2$) serves as the baseline and the speed-up for other values of $N_x=N_y=4,8, \ldots 4096,8192$  is shown in Fig. \ref{fig:box_size_scan}. There is an optimal value of the box size with respect to the triangle size in the range of $1<N_x/(2N_r)<10$. The speed-up for $N_x=256, 512, 1024$ are $35.5, 35.7, 36.3$; larger than those for other values of $N_x$.  For $N_x/(2N_r)\ll1$, each box contains a large number of triangles, as shown in Fig. \ref{fig:box_triangle} (left) and identifying the particle-triangle mapping consumes a significant amount of computing time. For $N_x=2$, the particle positioning in the charge deposition and field gathering can cost $>95\%$ of the total computing time. As $N_x/(2N_r)$ increases and becomes larger than 1, the particle positioning consumption is reduced and the cost of the charge deposition and the field gathering is comparable to the particle pusher. For $N_x/(2N_r)\gg1$, the memory cost for storing the box-triangle mapping increases but without significant CPU cost, and thus only slows down the simulation slightly. Even as $N_x$ changes from 1024 to 8192, the speed-up decreases from 36.3 to 32.7, by only around $10\%$.

\begin{figure*}\centering
	\includegraphics[width=0.3\textwidth]{\myfigure/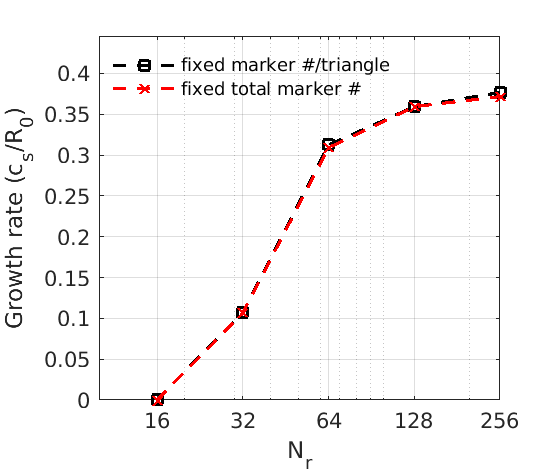}
	\includegraphics[width=0.3\textwidth]{\myfigure/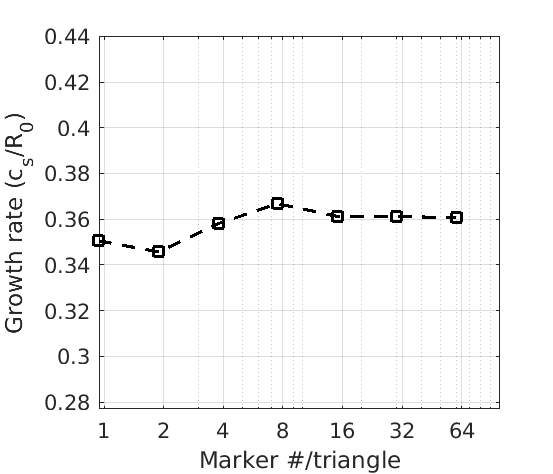}
	\includegraphics[width=0.3\textwidth]{\myfigure/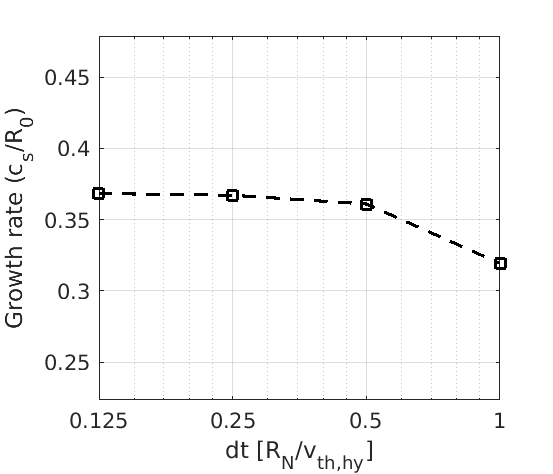}
	\caption{The convergence study for the growth rate of the ITG instability with respect to the grid number (left), marker number (middle) and time interval (right). The parameters for the base are $n=20$, $N_r=128$, $15$ markers per triangle, $dt=0.5$ and the scans are performed from the base case. In the left frame, the black line corresponds to the scan with 15 markers per triangle fixed. The red line with squares corresponds to the scan with 1.6 million markers fixed (15 markers per triangle for $N_r=128$).}
	\label{fig:convergence_marker_grid}
\end{figure*}

The convergence of the simulation results in terms of growth rate with respect to the radial grid number, the marker number per triangle and the time step size is studied. The convergence with respect to $N_r$, marker number $N$ and time step size $\Delta t$ is shown in Fig. \ref{fig:convergence_marker_grid}. For this $n=20$ mode, as shown in the left frame, from  $N_r=64$, the simulation starts to converge. The corresponding poloidal grid number per wave length $\approx7$. Note that the poloidal grid spacing is set to be close to the radial grid spacing. For the linear studies in this work, the mode structure is elongated along the radial direction and the minimum value of $N_r$ is determined by the grid number per wave length in poloidal direction, i.e., $N_\theta/m\approx2\pi N_r (r/a)/(nq)\gg1$.   In the middle frame, the results start to converge when the marker number per triangle $N/N_t>4$. In the right frame, the simulation starts to converge for $dt\leq0.5$ and becomes numerically unstable for $dt>1$.

\begin{figure}\centering
	\includegraphics[width=0.45\textwidth]{\myfigure/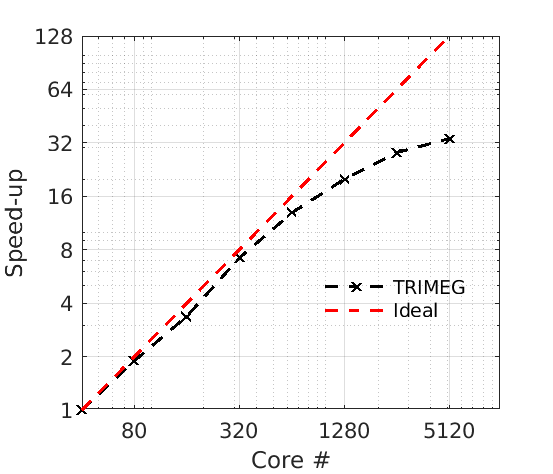}
	\caption{The speed-up of the simulation with respect to 40 cores versus the number of cores in TRIMEG simulations (black dashed line with crosses) and its comparison with the ideal scaling (red dashed line).
	}
	\label{fig:core_scaling}
\end{figure}

The parallel performance is tested for evaluating the  scaling properties. The speed-up for the simulation with $N=25.6$ million markers with different numbers of cores is analyzed and shown in Fig. \ref{fig:core_scaling}. Its comparison with the ideal scaling shows the good strong scaling for small to moderate core numbers (core number $\leq1280$). For even larger core numbers, the consumption of the field solver parallel communication can increase since the field solver is distributed over all cores. As a result, the deviation of the speed-up curve away from the ideal scaling becomes significant as the core number is larger than 1280.

\subsection{ITG simulation using Cyclone parameters}
\label{subsec:itg_cyclone}
\begin{figure}\centering 
	\includegraphics[width=0.45\textwidth]{\myfigure/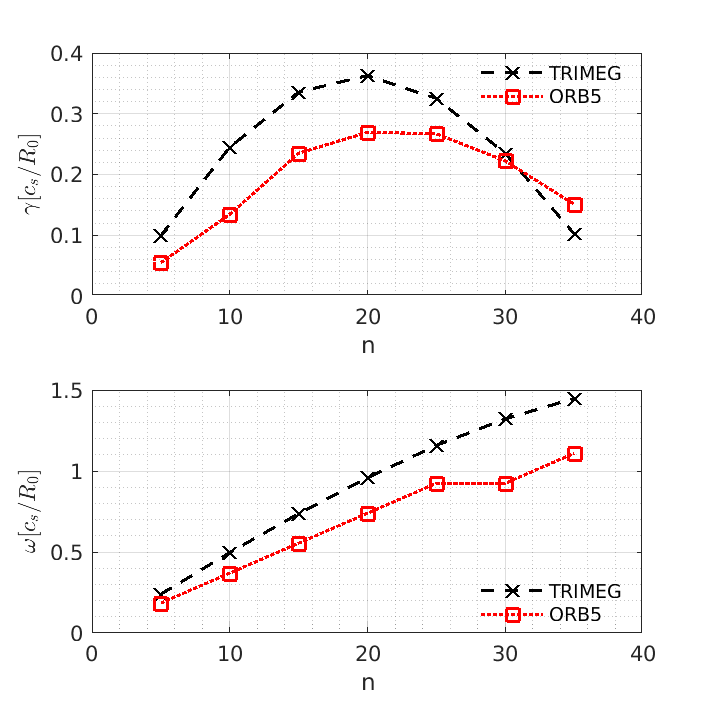} 
	\caption{Benchmark of the growth rate (upper) and the real frequency (bottom) for the Cyclone case. The ORB5 simulation results are from our previous work \cite{lu2017symmetry}.}
	\label{fig:eigenvalue_benchmark}
\end{figure}

Following the convergence and scaling studies in Section \ref{subsec:converge_scaling}, ITG mode simulations using the Cyclone case parameters are performed. The growth rate and the frequency are shown in Fig. \ref{fig:eigenvalue_benchmark} and are compared with the ORB5 results noticing $v_{ti}/R_N=\sqrt{2T_i/T_e}(R_0/R_N)(c_s/R_0)$, where $R_0$ is the major radius. The agreement between the TRIMEG results and the ORB5 results is reasonable, noting the simplifications in TRIMEG as discussed in Sec. \ref{subsec:parameters}. Nevertheless, considering the spatial scale separation between the equilibrium profile variation spatial scale $L_E$, the mode structure radial envelope width $L_A$ and the single poloidal harmonic width $1/(ndq/dr)$, i.e., $L_E\gg L_A \gg 1/(ndq/dr)$ \cite{lu2012theoretical}, the simulation from TRIMEG already captures the leading order solution. More comprehensive physics models will be implemented in the future.  

\subsection{ITG simulation using AUG equilibrium in the core plasma and in the whole plasma geometry}
\label{subsec:itg_aug}
\begin{figure*}\centering
	\includegraphics[width=0.32\textwidth]{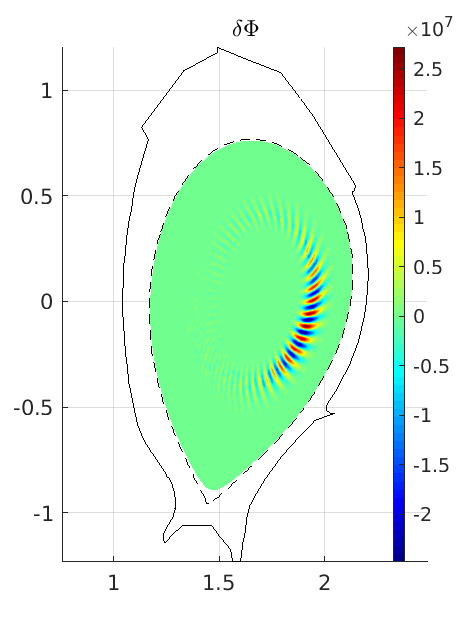}
	\includegraphics[width=0.32\textwidth]{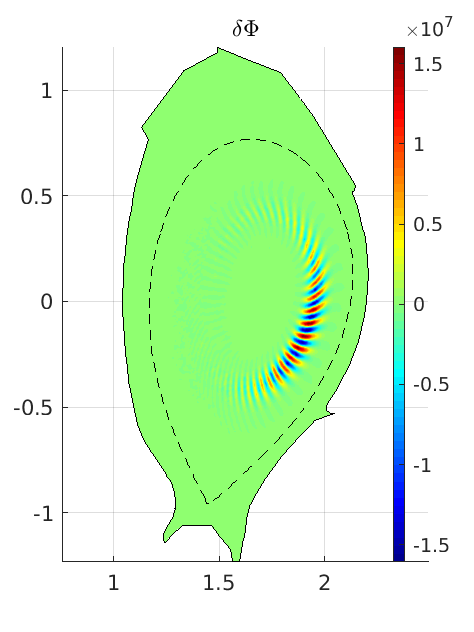}
	\includegraphics[width=0.32\textwidth]{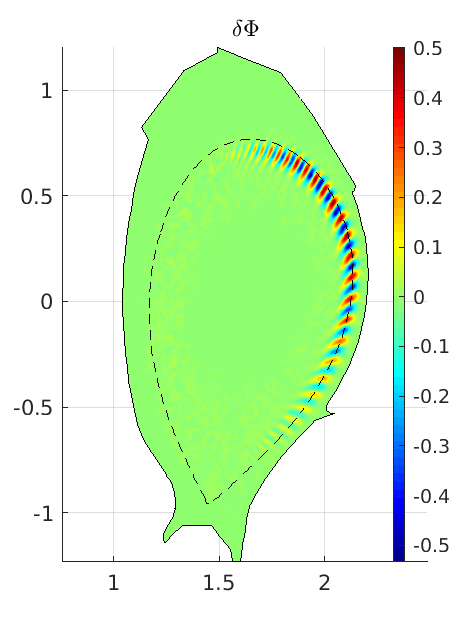}
	
	\caption{The 2D ITG mode structures ($n=20$) for the three cases defined in Table \ref{tab:aug3cases}. Case A (left), Case B (middle) and Case C (right). The dotted line indicates the separatrix.}
	\label{fig:aug_core_edge}
\end{figure*}

\begin{figure}\centering
	\includegraphics[width=0.4\textwidth]{\myfigure/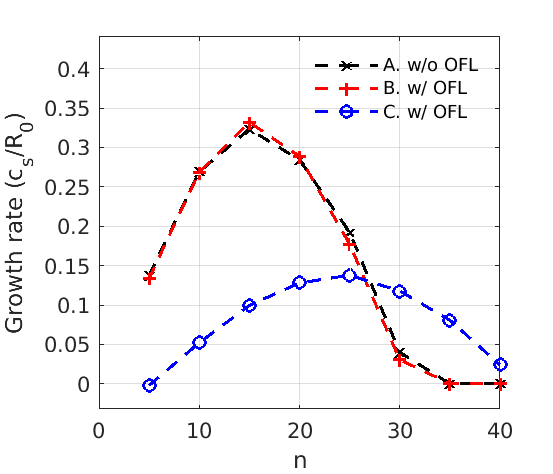}
	
	\caption{Growth rate versus toroidal mode number $n$ for the three cases defined in Table \ref{tab:aug3cases}. }
	\label{fig:aug_growth}
\end{figure}

In this section, we perform simulations of ITG modes using the AUG equilibrium described in Section \ref{subsec:parameters}. The main purpose is to demonstrate the capability of treating the realistic magnetic equilibrium from an experiment using TRIMEG. Three cases are defined in Table \ref{tab:aug3cases}. The simulations without and with the open field line region are performed and compared as shown in Fig. \ref{fig:aug_core_edge} and Fig. \ref{fig:aug_growth}. For Cases (A) and (B), the open field line plays a weak role on the core ITG mode due to the narrow envelope of the radial mode structure. As a result, the 2D mode structures and the growth rate are almost identical between these two cases. For Case C, the local variables at $\psi=\psi_b$ such as $q$, $\hat{s}$, and the minor radius that determine the mode growth rate are different from (A) and (B). As a result, the growth rate curve is shifted due to the change in the finite Larmor radius / finite orbit width effects ($k_\theta\rho_{th}\approx nq\rho_{th}/r$) and other effects.  While the convergence of the simulation for the whole plasma volume is achieved, a benchmark with other codes with the treatment of the whole plasma geometry will be studied in the future. Missing physics in the TRIMEG code includes, but is not limited to, the fully nonlinear collision operator \cite{hager2016fully}; a more realistic boundary condition such as a sheath boundary condition \cite{cohen2004sheath,boesl2019gyrokinetic}; the radial electric field consistent with neoclassical physics; and zonal flow physics. In addition, more comprehensive gyrokinetic/gyrofluid models for the edge also need to be considered  \cite{scott2006edge,qin2006general}. 

\begin{table}\begin{center}
		\begin{tabular}{c|c|c|c}
			\hline 
			Case & OFL & $\rho_{p,c}$ & $W_{p,c}$ \\ 
			\hline 
			A & w/o & 0.5 & 0.3  \\ 
			\hline 
			B & w/ & 0.5 & 0.3  \\ 
			\hline 
			C & w/ & 1.0 & 0.1  \\ 
			\hline 
		\end{tabular} 
	\end{center}\caption{Parameters for the three ITG simulations using the AUG equilibrium. OFL refers to the Open Field Line region. For all cases, $\kappa_n=2.23$, $\kappa_T=6.96$. }\label{tab:aug3cases}
\end{table}

\section{Conclusion} 
\label{sec:conclusion}
In this work, the TRIMEG code has been developed based on the mixed unstructured mesh based FEM-Fourier decomposition scheme and the intermediate grid for the particle position. The parallel scalability of charge deposition and field gathering has been achieved and strong scaling up to moderate core numbers has been demonstrated. The benchmark with ORB5 using the DIII-D Cyclone test case shows reasonable agreement in terms of growth rate and frequency. The capability of treating the whole plasma volume is demonstrated by using an AUG magnetic equilibrium with an X-point and analytical density and temperature profiles. Futher development of the TRIMEG code for specific physics studies such as the mode structure symmetry breaking \cite{lu2017symmetry,lu2018mode,lu2018kinetic,lu2019theoretical} and for more comprehensive simulations including the wave-particle and wave-wave nonlinearities and multiple species will be explored in the future. 

\section*{Acknowledgments}
Simulations were performed on MPCDF computing systems. Suggestions from ORB5, EUTERPE, HMGC, GTC and GTS groups, discussion with B.D. Scott, D. Coster, A. Bierwage and support by National Natural Science Foundation of China under Grant No. 11605186  are appreciated by ZL. ZL thanks X. Wang for naming the code TRIMEG.  Support by the ENR projects ``NAT'' and ``MET'' is acknowledged. This work has been carried out within the framework of the EUROfusion Consortium and has received funding from the Euratom research and training programme 2014-2018 and 2019-2020 under grant agreement No 633053. The views and opinions expressed herein do not necessarily reflect those of the European Commission. \\


\end{document}